\documentclass[%
 aip,
 amsmath,amssymb,
 reprint,%
]{revtex4-1}

\usepackage{graphicx}
\usepackage{dcolumn}
\usepackage{bm}

\usepackage[utf8]{inputenc}
\usepackage[T1]{fontenc}
\usepackage{mathptmx}
\usepackage{etoolbox}
\usepackage{epstopdf, epsfig, psfrag}
\usepackage{placeins,xcolor}
\usepackage{CJK}

\def\drawline#1#2{\raise 2.5pt\vbox{\hrule width #1pt height #2pt}}
\def\spacce#1{\hskip #1pt}
\def\solid{\drawline{24}{.8}\nobreak }

\def\bdash{\hbox{\spacce{1}\drawline{4}{.5}\spacce{1}}}
\def\bdashb{\hbox{\drawline{2}{.5}\spacce{1}}}
\def\dashed{\bdash\bdash\bdash\bdash\nobreak }

\def\dasheda{\bdashb\bdashb\bdashb\bdashb\bdashb\bdashb\bdashb\nobreak }

\def\trian{\raise 1.25pt\hbox{$\scriptscriptstyle\triangle$}\nobreak\ }
\def\trian{\raise 1.25pt\hbox{$\scriptscriptstyle\triangle$}\nobreak\ }

\def\square{${\vcenter{\hrule height .4pt
        \hbox{\vrule width .4pt height 5pt \kern 5pt
        \vrule width .4pt}
        \hrule height .4pt}}$\nobreak\ }

\def\rect{${\vcenter{\hrule height .4pt
        \hbox{\vrule width 0.4pt height 3.5pt \kern 14pt
        \vrule width 0.4pt}
        \hrule height .4pt}}$\nobreak}

\makeatletter
\def\@email#1#2{%
 \endgroup
 \patchcmd{\titleblock@produce}
  {\frontmatter@RRAPformat}
  {\frontmatter@RRAPformat{\produce@RRAP{*#1\href{mailto:#2}{#2}}}\frontmatter@RRAPformat}
  {}{}
}%
\makeatother
\begin{document}

\preprint{AIP/123-QED}

\title[Toward ultra-efficient high fidelity predictions of wind turbine wakes]{Toward ultra-efficient high fidelity predictions of wind turbine wakes: Augmenting the accuracy of engineering models with machine learning}
\author{C. Santoni}
\affiliation{ 
Department of Civil Engineering, Stony Brook University, Stony Brook, NY 11794, USA
}%
\author{D. Zhang \begin{CJK*}{UTF8}{gbsn} (张迪畅) \end{CJK*} }%
\affiliation{ 
Department of Computer Science, Stony Brook University, Stony Brook, NY 11794, USA
}%
\author{Z. Zhang\begin{CJK*}{UTF8}{gbsn} (张泽夏) \end{CJK*}}%
\affiliation{ 
Department of Civil Engineering, Stony Brook University, Stony Brook, NY 11794, USA
}%
\author{D. Samaras}%
\affiliation{ 
Department of Computer Science, Stony Brook University, Stony Brook, NY 11794, USA
}%
\author{F. Sotiropoulos}%
\affiliation{ 
Mechanical and Nuclear Engineering, Virginia Commonwealth University, Richmond, VA 23284, USA
}%
\author{A. Khosronejad$^*$}%
 \email{ali.khosronejad@stonybrook.edu.}
\affiliation{ 
Department of Civil Engineering, Stony Brook University, Stony Brook, NY 11794, USA
}%

\date{\today}

\begin{abstract}
This study proposes a novel machine learning (ML) methodology for the efficient and cost-effective prediction of high-fidelity three-dimensional velocity fields in the wake of utility-scale turbines. The model consists of an auto-encoder convolutional neural network with U-Net skipped connections, fine-tuned using high-fidelity data from large-eddy simulations (LES). The trained model takes the low-fidelity velocity field cost-effectively generated from the analytical engineering wake model as input and produces the high-fidelity velocity fields. The accuracy of the proposed ML model is demonstrated in a utility-scale wind farm for which datasets of wake flow fields were previously generated using LES under various wind speeds, wind directions, and yaw angles. Comparing the ML model results with those of LES, the ML model was shown to reduce the error in the prediction from 20\% obtained from the Gauss Curl Hybrid (GCH) model to less than 5\%. In addition, the ML model captured the non-symmetric wake deflection observed for opposing yaw angles for wake steering cases, demonstrating a greater accuracy than the GCH model. The computational cost of the ML model is on par with that of the analytical wake model while generating numerical outcomes nearly as accurate as those of the high-fidelity LES.
\end{abstract}

\maketitle

\section{Introduction}
The increasing demand for a clean and sustainable energy source has led to a surge in the number and size of wind turbines and wind farms. 
However, wake interference may decrease the efficiency of these arrays.
Turbine wakes are accountable for power losses of up to $10\%$ to $20\%$~\cite{Barthelmie2007, Barthelmie2009}. 
Furthermore, through supervisory control and data acquisition (SCADA) and light detection and ranging (LiDAR) measurements of a wind farm in Texas, El-Asha et al.~\cite{el-asha} observed significant power losses on specific turbines between 60\% to 80\% due to wake shadowing. 
To mitigate the impact caused by wake interaction and maximize the energy production of a wind farm, it has been suggested to deliberately adjust the yaw of turbines, aiming to redirect their wakes away from the downstream turbines and thus minimizing turbine wake interactions~\cite{ Medici2006, Jimenez2010,  wagenaar2012, Fleming2014, HerbertAcero2014, Gebraad2016,  Boersma2017}.
Nonetheless, to effectively design active wake controllers, it is essential to comprehend wake dynamics and how a turbine wake interacts with downstream turbines~\cite{kang2022}.
Although high-fidelity numerical modeling could provide enough physical details, their computational cost can be prohibitive for addressing optimization problems~\cite{liu2022}.
Hence, it has been imperative to develop simplified wake models that describe the dynamics of the wake with sufficient details, all while maintaining computational efficiency.

Earlier models for wind farm optimization were developed for \textit{infinite} turbine arrays by considering turbines as an increased roughness in the atmospheric boundary layer~\cite{Templin1974, Newman1977, Frandsen1992}.
This concept was further applied by Newman~\cite{Newman1977}, who approximated the optimal turbine spacing to be approximately $28D$ (where $D$ is the turbine diameter) to achieve a power loss of less than $10\%$.
While such over-simplified models offer a straightforward representation of a wind farm in the atmospheric boundary layer, their applicability to finite-size wind farms might compromise their accuracy. 
In response to this challenge, Jensen~\cite{Jensen1983} introduced an analytical turbine wake model to represent each turbine. 
The model is characterized by a uniform velocity profile behind the rotor and a linear wake expansion that accounts for the energy entrainment. 
This approach treats the interaction between wakes from multiple turbines as a superposition of the momentum deficit.
Further improvement to Jensen's wake model was made by treating the wake interaction as the supper position of the energy deficits~\cite{Katic1986}, which results in the equilibrium of the total velocity deficit after 3 to 4 turbine rows.

As recognized by Jensen~\cite{Jensen1983} and Kati{\'c} et al.~\cite{Katic1986} and shown by experimental measurements~\cite{Chamorro2009, Chamorro2010}, the top-hat description of the velocity in their wake model is not realistic and that a Gaussian distribution could provide a better alternative.
Moreover, Jensen's model is primarily based on mass conservation in the turbine wake without explicitly addressing the momentum conservation~\cite{Bastankhah2014}. 
To address these shortcomings, Bastankhah and Port{\'{e}}-Agel~\cite{Bastankhah2014} applied momentum and mass conservation and assumed a self-similar Gaussian velocity deficit in the wake of the turbine to derive a new analytical model. 
The Gaussian wake model reasonably agreed with high-fidelity numerical simulations of the actuator disk model with rotation (ADM-R) of Wu and Port{\'{e}}-Agel~\cite{wu2010, Wu2012} and wind tunnel measurements of Chamorro and Port{\'{e}}-Agel~\cite{Chamorro2010}.
Moreover, Niayifar and Port{\'{e}}-Agel~\cite{Niayifar2016} extended the application of the Gaussian wake model for the prediction of the velocity and power production of large wind farms. 
In this model, the wake interaction between the turbines is considered the superposition of the velocity at the turbine rotor and the wake velocity. 
In addition, Bastankhah and Port{\'{e}}-Agel~\cite{Bastankhah2014} proposed an empirical relation between wake growth rate ($k^*$) and the turbulence intensity, which increases due to the presence of the turbines. 

Due to the interest in applying wind farm yaw control for wake steering, existing engineering wake models were modified to predict the velocity deficit in such cases. 
For example, Jim{\'e}nez et al.~\cite{Jimenez2010} developed a top-hat skew wake model similar to Jensen~\cite{Jensen1983} and Kati{\'c} et al.~\cite{Katic1986}.
Moreover, Bastankhah and Port{\'{e}}-Agel~\cite{Bastankhah2016} expanded their Gaussian wake model to account for the wake displacement due to the yaw misalignment.
They showed their model to predict the momentum deficit and the wake deflection with good accuracy, compared to wind tunnel measurements of a single turbine.
However, comparisons of the Gaussian wake model against large-eddy simulation (LES) results of the ADM-R have shown that the pair of counter-rotating vortices (CVP) that are induced by the yaw misalignment---and neglected by the wake model---causes a nonsymmetrical deflection of the wake to oppose yaw angles and secondary steering of downstream turbine wakes~\cite{fleming2018a}. 
Furthermore, Ciri et al.~\cite{Ciri2018} showed that the rotor size influences the length and time scale of the wake vortical structures, thus retaining the initial orientation induced by the yaw before they break down. 
This implies that larger turbines cause a larger deflection of the wake.
To address the challenge presented by the counter-rotating vortices, Mart{\'i}nez-Tossas et al.~\cite{Martinez-Tossas2019, martinez-tossas2021} considered the curling of the wake caused by CVP as the superposition of smaller Lamb-Oseen vortices that act on the base flow.
The curled wake model showed good agreement with LES and experimental measurements. 
Although the Reynolds Averaged Navier-Stokes (RANS) implementation is more computationally efficient than LES at the cost of fidelity, it increases the computational cost significantly compared to the simple engineering wake models. 
To address the secondary steering, King et al.~\cite{King2021} introduce the curled wake model of Mart{\'i}nez-Tossas et al.~\cite{Martinez-Tossas2019, martinez-tossas2021} to compute the cross-wind velocity in the wake of the turbines to the Gaussian wake model and compute an effective yaw angle for the computation of the wake of downstream turbines.
Comparing their Gaussian-curl hybrid (GCH) model against LES using the Simulator for Wind Farm Applications (SOWFA) of small and large-scale wind farms, King et al.~\cite{King2021} showed better agreement than the standard Gaussian wake model.

High-fidelity numerical simulations and LiDAR measurements can aid data-driven reduced-order models in accurately characterizing utility-scale turbine wakes, overcoming challenges that analytical wake models may face.
For example, Zhang et al.~\cite{Zhang2021} developed an auto-encoder convolutional neural network (ACNN) to predict turbine wake velocity with only a few snapshots of the instantaneous velocity field.
The ACNN model was trained using the LES results of the Sandia National Laboratories Scaled Wind Farm Technology (SWiFT) site. 
Their wake flow predictions with the ACNN model for the aligned turbines showed great agreement with their simulation results and reduced the computational cost by $88\%$. 
In a similar attempt, Renganathan et al.~\cite{ashwinrenganathan2022} developed a data-driven machine learning (ML) model from LiDAR measurements of a wind farm in Texas.
The model initially consisted of an encoder-decoder convolutional neural network (CNN) trained with LiDAR measurements to obtain a low-dimensional latent space.
Subsequently, the encoder of the CNN was replaced by a multi-layer perceptron and a Gaussian Predictive (GP) model to reproduce the latent space from their input parameters and reproduce their measurements using the encoder of their CNN.
Ti et al.~\cite{ti2020} used the data obtained from a RANS-based model to train multiple single-layer neural networks to predict the wind speed and turbulence kinetic energy (TKE) in the turbine wake based on the hub height wind speed and turbulence intensity.
Moreover, their neural network outperformed the Jensen and Gaussian wake models by concatenating their machine learning model. 
It could predict the velocity field and the TKE for multiple rows of turbines more accurately than the RANS model~\cite{ti2020}. 
It is important to mention that machine learning model applications mainly focus on turbine power production~\cite{stanfel2020, gao2021}. 
However, Santoni et al.~\cite{Santoni2023} expanded the work of Zhang et al.~\cite{Zhang2021} and developed an ACNN model to predict the velocity and TKE in the wake of the turbines with a few snapshots of the instantaneous velocity obtained from LES.
The results obtained from the ACNN model were shown to predict the wake velocity and deflection well with good accuracy compared to those of the LES model.
More specifically, their ACNN model could accurately predict the secondary wake steering on the downstream turbine in the SWiFT site.
While their ACNN model decreased the computational cost of the high-fidelity simulation by $85\%$, they still required $10\,000$ CPU hours to generate the instantaneous velocity field snapshots using LES. 
Therefore, the application of their ACNN model for yaw control optimization was limited by high computational cost.

To tackle the aforementioned challenges, we have developed a data-driven machine-learning-based wake model for ultra-efficient inference of the high-fidelity velocity field in the wake of the turbines. 
This new model is similar to that developed by Zhang et al.~\cite{Zhang2021} and Santoni et al.~\cite{Santoni2023}. 
However, the new ACNN model effectively narrows the computational cost gap between an engineering wake model and LES to achieve a high-fidelity field, as its input parameters consist of the velocity field obtained from an analytical wake model.
The ML model was trained and validated against LES, performed with the Virtual Flow Simulator (VFS), where the turbines were parameterized using the actuator surface model for the blades and nacelle. 
Moreover, we have shown that the ACNN model increases the accuracy of the engineering wake model in comparison to the high-fidelity simulations and that the developed ML model decreased the computational cost of the high-fidelity simulation by $99\%$. 
The computational cost of this ML model is significantly reduced by replacing the LES inputs with the inputs from the low-fidelity wake model and, therefore, addressing the shortcomings of the ACNN model reported by Zhang et al.~\cite{Zhang2021} and Santoni et al.~\cite{Santoni2023}. 
As a result, the developed ML algorithm allows for effective yaw control and control co-design optimization of wind farms.

The paper is organized as follows. A description of the Virtual Flow Simulator (VFS) Wind is given in Section~\ref{sec:GovEq}.
In addition, a brief description of the engineering wake model and the Machine Learning (ML) model architecture are given in Section~\ref{sec:GCH} and Section~\ref{sec:ACNN}, respectively.
Details of the LESs are given in Section~\ref{sec:CD}.
Comparison of the numerical results obtained from the LES, the engineering wake model, and the ML model for various wind directions and yaw angles are given in Section~\ref{sec:RandA}.
Summary and final remarks can be found in Section~\ref{sec:summary}.

\section{Methodology} \label{sec:methodology}

\subsection{Equations of flow motion}\label{sec:GovEq}
Large-eddy simulations were performed using the Virtual Flow Simulator-Wind (VFS-Wind) model that solves the filtered Navier-Stokes equations in a generalized curvilinear coordinates system given by
\begin{subequations}\label{eq:N-S}
\begin{eqnarray}
    \frac{\partial U^i}{\partial t} &=& \xi^i_l \left[ -\frac{\partial U^j u_l}{\partial \xi^j} + \frac{1}{Re} \frac{ \partial }{\partial \xi^j}\left(\frac{g^{jk}}{J} \frac{\partial u_l}{\partial \xi^k}\right ) \right . \nonumber \\
    & & \left .- \frac{\partial }{\partial \xi^j}  \left(\frac{\xi^j_l p}{J} \right)   - \frac{\partial \tau_{lj} }{\partial \xi^j} +f_l \right ], \\
   J \frac{\partial U^i}{\partial \xi^i} &=& 0,
\end{eqnarray}
\end{subequations}

where $p$ is the pressure and $u_l$ is the filtered Cartesian velocity component along the $l$-direction.
The contravariant volume flux ($U_i$) is given by $U_i = (\xi^i_m/J)u_m$, where $\xi^i_m$ and $J$ is the Jacobian of the curvilinear transformation and its determinant, respectively.
The contravariant tensor components are given by $g_{ik} = \xi^i_l \xi ^k_l$.
The Reynolds number is defined as $Re= U_\infty D/\nu$, where $U_\infty$ is the wind velocity far from the bottom surface, $D$ is the wind turbine rotor diameter, and $\nu$ is the kinematic viscosity.
The sub-grid stresses are modeled using the dynamic eddy viscosity model~\cite{germano1991}.
The external forces per unit volume, $f_l$, were computed using the actuator surface model~\cite{yang2018} for the wind turbine blades and nacelle. 
The momentum equations were discretized in space using the central-differences scheme and advanced in time using the Crank-Nicolson scheme.
The fractional step method was used to project the resulting non-solenoidal velocity field into a solenoidal space~\cite{kim1985}.

The wind turbine rotor and nacelle have been parameterized using the actuator surface model (ASM) developed by Yang and Sotiropoulos~\cite{yang2018}.
The ASM models the turbine by computing the lift and drag coefficient based on the blade element theory~\cite{froude1878} over a two-dimensional unstructured grid and distributed over the flow grid using a smoothed four-point cosine function proposed by Yang et al.~\cite{yang2009}. 
The lift and drag coefficients were corrected to account for the three-dimensional (3D) and rotational effects using Du et al.~\cite{du1998} stall delay model.
In addition, tip losses due to the formation of the blade tip vortex were considered by applying the tip loss correction factor of Shen et al.~\cite{shen2005} to the computed forces. 

The rotor angular velocity ($\omega$) was computed from the balance of angular momentum equation given by
\begin{equation}
    I \frac{d \omega}{dt} = \tau_{a} + \tau_{g},
\end{equation}
where $I$ is the rotational inertia, $\tau_{a}$ and $\tau_{g}$ are the aerodynamic and the generator torque, respectively.
At wind speeds below rated by the manufacturer, the power production is maximized by regulating the angular velocity of the rotor by setting the generator torque at~\cite{burton2011}
\begin{equation}
    \tau_g = \frac{ \pi \rho R^5 C_{p,max}}{2 \lambda_{opt}^3 G^3} \omega_g^2,
\end{equation}
where $C_{P,max}$ is the maximum power coefficient at the optimal tip-speed ratio $\lambda_{opt}$, $R$ is the rotor radius, $G$ is the gearbox ratio, and $\omega_g$ is the generator angular velocity.
Details of the controller can be found in Santoni et al.~\cite{Santoni2023b}.

\subsection{Analytical wake model: Gauss-Curl Hybrid}\label{sec:GCH}
The wake model carried out for the input of the developed machine learning model is the Gauss-Curl Hybrid model ~\cite{King2021}, as implemented in FLORIS (v3.4)~\cite{Fleming2020}.
The GCH combines the Gaussian wake model developed by Bastankhah and Port{\'{e}}-Agel~\cite{Bastankhah2014, Bastankhah2016}, and Niayifar and Port{\'{e}}-Agel~\cite{Niayifar2016} with curl model developed by Mart{\'i}nez-Tossas et al.~\cite{Martinez-Tossas2019}.
The curl model estimates the spanwise velocity due to the yaw-induced counter-rotating vortices.
These spanwise velocity components from the curl model are incorporated to consider the added recovery due to yaw misalignment and secondary wake steering of a downstream turbine.
The wake interactions are taken into account by the sum of the energy deficits of Kati{\'c} et al.~\cite{Katic1986}, also known as the sum of the square method.
For brevity, the reader is referred to King et al.~\cite{King2021} for a full description of the GCH model.
Input parameters for the GCH model were taken from the example input file of the FLORIS (v3.4) package provided. 
It should be noted that using these input parameters naively may not yield the best results, and tuning may improve the results obtained from the GCH model. 
These were taken as is to demonstrate the capabilities of the machine learning model.
However, the ambient turbulence intensity (TI) was set to 15\%, corresponding to that of the precursor simulation of the LES~\cite{Zhang2021}. 
The power and thrust coefficient of the Vestas V27 turbine were obtained from an in-house BEM model.

\subsection{Autoencoder convolutional neural network}\label{sec:ACNN}
\begin{figure*}[tb]
 \begin{center}
  \psfrag{alab1}[][][0.7]{142}
  \psfrag{alab2}[][][0.7]{64}
  \psfrag{alab3}[][][0.7]{128}
  \psfrag{alab4}[][][0.7]{256}
  \psfrag{alab5}[][][0.7]{512}
  \psfrag{alab6}[][][0.7]{1024}
  \psfrag{alab8}[][][0.7]{32}
  \psfrag{blab1}[][][0.7]{12}
  \psfrag{blab2}[][][0.7]{32}
  \psfrag{blab3}[][][0.7]{32}
  \psfrag{text1}[bl][bl][0.8]{Max Pooling ($2\times2$)}
  \psfrag{text2}[bl][bl][0.8]{De-convolution (Upsampling)}
  \psfrag{text3}[bl][bl][0.8]{Convolution ($3\times3$) + ReLU}
  \psfrag{text4}[bl][bl][0.8]{Convolution ($1\times1$)}
  \psfrag{text5}[bl][bl][0.8]{Copy}
  \psfrag{text6}[bl][bl][0.8]{Linear + ReLU}
  \includegraphics[width=\textwidth]{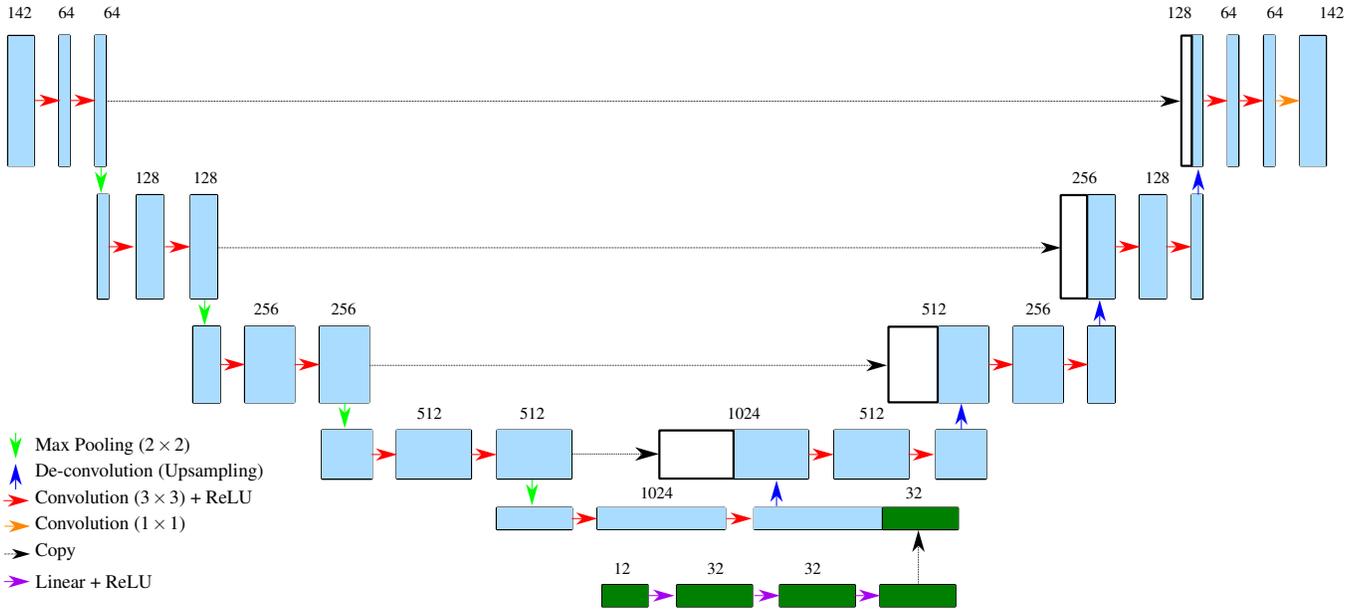}
  \caption{Autoencoder convolutional and fully-connected neural network architecture. Arrows indicate the sub-layer process. The number of channels/neurons is given by the number next to the latent space representation boxes.
  }
  \label{fig:Arch}
 \end{center}
\end{figure*}
The machine learning (ML) model implemented in this work consists of an autoencoder convolutional network (ACNN) architecture inherited from U-Net~\cite{ronneberger2015}.
A schematic of the implemented ACNN model architecture is shown in Figure~\ref{fig:Arch}.
A fully connected neural network (FCN), which consists of three linear layers followed by a rectified linear units (ReLU) activation function~\cite{maas2013}, processes the model input parameters, which are then concatenated to the latent sub-space obtained from the CNN encoder, as shown in Figure~\ref{fig:Arch}.
The encoder of the CCN, which processes the velocity field obtained from the GCH model, comprises five layers, each containing two convolutions sub-layer and a max pooling sub-layer.
A ReLU activation function immediately follows each convolution sublayer.
The max pooling sub-layer decreases the sub-space dimension, which is then passed to the next layer to extract higher-level semantic features.
After the latent subspace obtained from the FCNN is concatenated with that obtained from the encoder, the resulting subspace is passed to the decoder.
The decoder consists of 4 layers, each with a deconvolution (up-sampling) sub-layer followed by two convolution sub-layers.
In addition, the up-sampled feature layer is concatenated with the intermediate features obtained from the encoder.
This is known as the skipped connection technique introduced by the U-Net~\cite{ronneberger2015} that enables high-fidelity predictions.
Finally, a $1\times 1$ convolutional layer accepts features from the last decoder stack and produces final predictions.

\subsubsection{Training and inference}
\begin{figure}[tb]
 \begin{center}
  \psfrag{alab}[l][][0.9]{a)   Training}
  \psfrag{blab}[l][][0.9]{b)   Inference}
  \psfrag{vfswind}[][][0.7]{LES}
  \psfrag{lab3}[][][0.7]{Input}
  \psfrag{lab1}[][][0.7]{GCH}
  \psfrag{lab2}[][][0.7]{FCNN}
  \psfrag{inp1}[][l][0.7]{Low Fidelity}
  \psfrag{cnn1}[][][0.7]{Encoder}
  \psfrag{cnn2}[][][0.7]{Decoder}
  \psfrag{acnn}[][][0.7]{ACNN}
  \psfrag{output}[][][0.7]{High Fidelity}
  \includegraphics[width=0.48\textwidth]{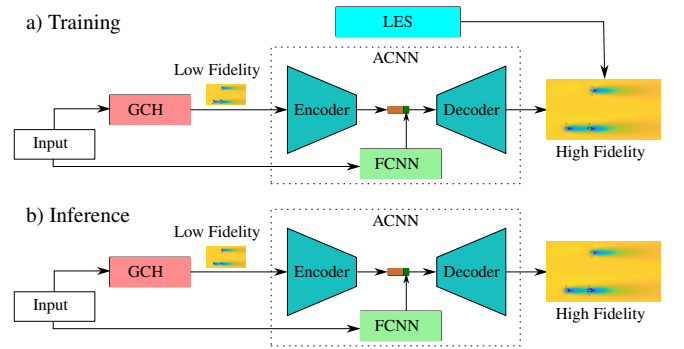}
  \caption{Schematic of the machine learning model (a) training and (b) inference procedure. The input parameters are passed to a Fully Connected Neural Network (FCNN) and the Gauss Curl Hybrid (GCH) model. The low-fidelity velocity field obtained from the GCH model is passed to the Auto-encoder Convolutional Neural Network (ACNN). The output of the FCNN is concatenated to the latent subspace created by the decoder part of the ACNN. The output of the ACNN corresponds to a high-fidelity approximation of the velocity field.
  }
  \label{fig:MLSch}
 \end{center}
\end{figure}
The input to the ML model consists of a low-fidelity description of the wake obtained from the engineering wake model and the input parameters, such as the location of the turbines, yaw angle, wind speed, turbulence intensity, and wind shear coefficient.  
During the training, backward propagation was used to minimize the error between the time-averaged velocity fields obtained from the LES and the model (see Figure~\ref{fig:MLSch}a).
The network was trained for $50\,000$ epochs to minimize the mean square error between network outputs and LES simulations using the Adam optimizer~\cite{Kingma2015} with a learning rate of $10^{-4}$. 
The training set fits in a single batch, so each epoch represents a network update. 
A dropout~\cite{Srivstava2014} rate of $0.2$ was added to prevent overfitting.
The training was performed on a single RTX A6000 GPU and took around $48$ hours of wall-clock time.
The results obtained from the ML model at inference (see Figure~\ref{fig:MLSch}b), were compared against the numerical results from the LES for validation.

\section{Test case description and computational details}\label{sec:CD}
\begin{figure*}[tb]
 \begin{center}
  \psfrag{alab}[][][1.0]{ }
  \psfrag{blab}[][][1.0]{ }
  \psfrag{xlab}[][][0.8]{$x/D$}
  \psfrag{ylab}[][][0.8]{$y/D$}
  \psfrag{zlab}[][][0.8]{$z/D$}
  \psfrag{a1lab}[][][0.7]{$3D$}
  \psfrag{a2lab}[][][0.7]{$7.8D$}
  \psfrag{a3lab}[][][0.7]{$1.1D$}
  \psfrag{a4lab}[][][0.7]{$5D$}
  \psfrag{a5lab}[][][0.7]{$8D$}
  \psfrag{nlab}[][][0.7]{North}
  \psfrag{slab}[][][0.7]{South}
  \psfrag{T1}[][][0.7]{T1}
  \psfrag{T2}[][][0.7]{T2}
  \psfrag{T3}[][][0.7]{T3}
  \includegraphics[width=\textwidth]{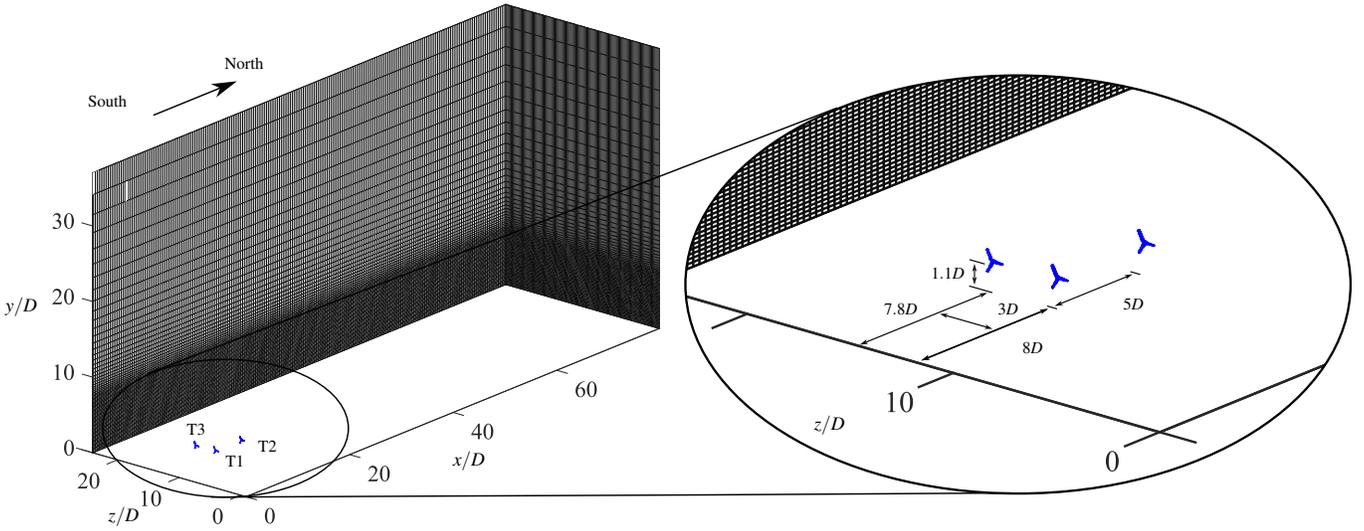}
  \caption{Geometrical configuration of the computational domain for the south wind direction. The Eulerian computational grid system ({\color{gray} \solid}) is shown for every other computational cell.}
  \label{fig:CB}
 \end{center}
\end{figure*}
A series of large eddy simulations were carried out to obtain high-fidelity data of the SWiFT facility wake flow field under various wind and yaw conditions. 
The SWiFT facility in Lubbock, Texas, consists of three Vestas V27 experimental turbines with a nameplate capacity of 225 kW at a rated wind speed of $14$ m/s. 
The turbines have a rotor diameter of $D=27$ m at a hub height of $32.1$ m. 
Details of the SWiFT facility can be found in Berg et al.~\cite{Berg2014}.

The turbines are placed on flat terrain with a dimension of $80 D \times 24.9D \times 37D$ along the streamwise ($x$), spanwise ($z$), and vertical ($y$) directions, respectively. 
The number of computational grid points is $451 \times 143 \times 281$ in the streamwise, spanwise, and vertical directions, respectively, which results in a resolution of $\Delta x/D = 0.16$ and $\Delta z/D =0.08$ in streamwise and spanwise directions, respectively. 
In the vertical direction, the grid has a uniform resolution of $\Delta y/D = 0.08$ up to a height of $y/D=4.4$, above which the grid is stretched to the top of the domain. See Figure~\ref{fig:CB} for the wind direction incoming from the south. 

A free slip boundary condition is imposed at the top boundary and periodic boundary condition along the spanwise direction.
The logarithmic law of the wall is applied to the bottom boundary and is given by
\begin{equation}
    U = \frac{U_*}{\kappa}\ln{\frac{y}{y_0}},
\end{equation}
where $U_*$ is the friction velocity, $\kappa$ is the von Karman constant, and $y_0$ is the roughness length ($y_0=0.004D$).
An inflow-outflow boundary condition is imposed along the streamwise direction. 
A precursor simulation with periodic boundary conditions was performed to obtain a fully developed neutral atmospheric boundary layer. 
The transient of the precursor simulation was discarded until the total kinetic energy reached a quasi-steady state. 
Thereafter, a cross-sectional plane from the instantaneous velocity fields of the precursor simulation was fed at the inlet of the wind turbine simulation. 

To generate different wake interaction conditions between the turbines, large eddy simulations were performed for four wind directions:  $30^\circ$ (northeast), $180^\circ$ (south), $210^\circ$ (southwest), and $266^\circ$ (west).
In addition, five different wind speeds were considered for each wind direction: $U_{hub} = 4.8$, $6.1$, $7.5$, $8.8$, and $10.2$ m/s.
For the wake steering cases, the computational box was decreased to $40.2D \times 12.4D \times 18.5D$ along the streamwise, spanwise, and wall-normal directions, respectively.
As a result, the resolution is $\Delta x/D = 0.08$, $\Delta z/D =0.04$ and $\Delta y/D =0.04$ at the rotor region.
The numerical simulations were performed for the South wind direction configuration for $U_{hub} = 5.4$, $7.0$, $8.5$, $10.1$, and $11.6$ m/s. For each wind speed, fixed yaw angles were imposed at the turbine $T1$, $\gamma_{T1} = -20^\circ$, $-15^\circ$, $-10^\circ$, $0^\circ$, $10^\circ$, $15^\circ$, and $20^\circ$ (see Figure~\ref{fig:CB}).

The numerical results obtained from the LES of the Vestas V25 wind turbine were validated against nacelle-mounted LiDAR measurements of Herges and Keyantuo~\cite{herges2019} at the SWiFT facility and reported in Zhang et al.~\cite{Zhang2021}.
In addition, a grid sensitivity analysis of the actuator surface model can be found in Santoni et al.~\cite{santoni2024}. 

\section{Results and discussions}\label{sec:RandA}
\subsection{Wind direction}
\begin{table}[b]
\caption{Root mean square error percentage of the velocity field prediction of the GCH and ML model compared to the LES. Empty spaces (---) indicate the training case.}
\centering
\begin{tabular}{lcccccccc}
\hline \hline
\multicolumn{1}{c}{$U_{hub}$ [m/s]} &     \multicolumn{8}{c}{Wind direction}    \\
\hline
\multicolumn{1}{c}{}& \multicolumn{2}{c}{Northeast} & \multicolumn{2}{c}{South} & \multicolumn{2}{c}{Southwest} & \multicolumn{2}{c}{West} \\
      & GCH    & ML      & GCH     & ML         & GCH      & ML   & GCH    & ML        \\
4.8   & $8\%$   & $2\%$    & ---     & ---      & $7\%$  & $1\%$  & $7\%$  & $1\%$      \\
6.1   & $7\%$   & $1\%$    & $6\%$    & $1\%$   & $8\%$  & $1\%$  & $8\%$  & $2\%$      \\
7.5   & $7\%$   & $1\%$    & $6\%$    & $1\%$   & ---    & ---    & $7\%$  & $1\%$      \\
8.8   & $7\%$   & $1\%$    & $7\%$    & $1\%$   & $7\%$  & $1\%$  & $7\%$  & $1\%$      \\
10.2  & $7\%$   & $1\%$    & $7\%$    & $2\%$   & $7\%$  & $1\%$  & ---    & ---       \\
\hline \hline
\end{tabular}
\label{table:rmse_wdir}
\end{table}
\begin{figure*}[tb]
 \begin{center}
  \psfrag{alab}[][][0.8]{a)}
  \psfrag{blab}[][][0.8]{b)}
  \psfrag{clab}[][][0.8]{c) }
  \psfrag{xlab}[][][1.0]{$x/D$}
  \psfrag{ylab}[][][1.0]{$z/D$}
  \psfrag{t1}[][][0.5]{T2}
  \psfrag{t2}[][][0.5]{T3}
  \psfrag{t3}[][][0.5]{T1}
  \psfrag{cblab}[][][1.0]{$\overline{U}/U_{hub}$}
  \includegraphics[width=\textwidth]{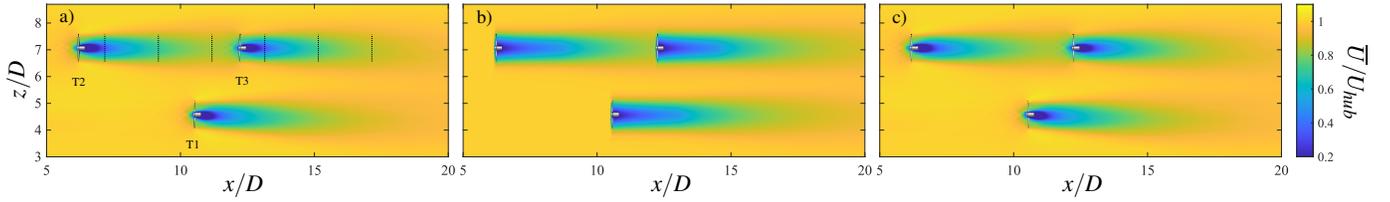}
  \caption{Contours of time-averaged velocity ($\overline{U}$) at a hub height plane for the North-East wind direction and hub height wind speed of $U_{hub}=8.8$ m/s at the SWiFT turbines facility; (a) LES, (b) GCH model, and (c) ML model. The streamwise and spanwise directions are given by $x/D$ and $y/D$, respectively, where $D$ is the turbine rotor diameter.
  }
  \label{fig:VelNE}
 \end{center}
\end{figure*}
The LES data of the wake flow for three cases of wind conditions, i.e., the south $U_{hub} = 4.8$ m/s, southwest $U_{hub} = 7.5$ m/s and west $U_{hub} = 10.2$ m/s, were selected to train the ML model. 
After completing the training of the ML model, it was employed to predict the mean wake flow field of the wind farm for other wind conditions.   
The root mean square error (RMSE) or the mean wind speed relative to the high-fidelity simulation results of LES (shown in Table~\ref{table:rmse_wdir}) were computed as 
\begin{equation}
    RMSE = \sqrt{\frac{1}{N} \sum_{i=1}^N (\overline{U}_{model}-\overline{U}_{LES})_i^2},
\end{equation}
where $\overline{U}$ is the time-averaged velocity at grid point $i$ and $N$ is the total number for grid points. 
It should be noted that the ML model was not exposed to the LES results of these test cases during its training process. 
As seen in this table, the RMSE of the GCH model ranges between $6\%$ to $8\%$, while the RMS of the ML model is between $1\%$ to $2\%$ of the LES results, showing that the errors of the ML model are $4$ to $7$ times smaller than those of the GCH model.
Interestingly, the ML model does not present any evident increase in the error for the wind speeds or wind direction cases not included in the model training.

For the sake of brevity, the discussion of the mean velocity field will be limited to the cases of northeast wind direction at $U_{hub}=8.8$ m/s and the west wind direction at $U_{hub}=6.1$ m/s.
The former pertains to a wind direction and wind speed outside of the machine learning training dataset, while the latter corresponds to a wind direction where the interacting wakes of turbines are closest, involving a wind speed not covered in the training data.
Contours of the time-averaged wind speed of the North-East case are shown in Figure~\ref{fig:VelNE}.
As observed from the contours obtained from LES (Figure~\ref{fig:VelNE}a), the wake of the turbines marks a pronounced momentum deficit in the center of the wake caused by the nacelle that recovers rather quickly further downstream.
The GCH model (Figure~\ref{fig:VelNE}b) also shows a large momentum deficit in the center of the wake that -- relative to the LES results -- persists further downstream.
In other words, the GCH model shows a slower wake recovery than the high-fidelity simulation.
The ML model (Figure~\ref{fig:VelNE}c) improves upon the GCH model, which served as its input and more closely resembles the turbine wake of the LES.

\begin{figure}[tb]
 \begin{center}
  \psfrag{xlab}[][][1.0]{$\overline{U}/U_{hub}$}
  \psfrag{ylab}[][][1.0]{$(z-z_{T2})/D$}
  \psfrag{alab}[][][1.0]{a)}
  \psfrag{blab}[][][1.0]{b)}
  \psfrag{clab}[][][1.0]{c)}
  \psfrag{dlab}[][][1.0]{d)}
  \psfrag{elab}[][][1.0]{e)}
  \psfrag{elab2}[][][1.0]{f)}  
  \includegraphics[width=0.48\textwidth]{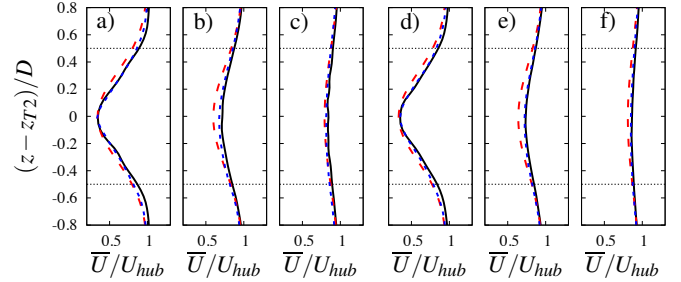}
  \caption{Time-averaged velocity ($\overline{U}$) profiles along the spanwise direction ($x$) at a hub-height plane of the North-East wind direction $U_{hub}=8.8$ m/s case at (a) $1D$, (b) $3D$, (c) $5D$ downstream from turbine T2 and (d) $1D$, (e) $3D$, (f) $5D$ downstream from turbine $T3$; (\solid) LES, ({\color{red}\dashed}) GCH model and ({\color{blue}\dasheda}) ML model. The profiles were centered along the spanwise direction with the location of turbine $T2$ ($z_{T2}$).
  }
  \label{fig:ProfNEP}
 \end{center}
\end{figure}

Velocity profiles along the spanwise direction of the two aligned turbines ($T2$ and $T3$) are shown in Figure~\ref{fig:ProfNEP}.
Although the results obtained from the LES and the GCH model at $1D$ from turbine $T2$ (Figure~\ref{fig:ProfNEP}a) highlight a reasonably good agreement at the center wake, near the edges ($|z-z_{T2}/D|>0.2$) it is observed that the GCH model over-estimate the momentum deficit.
Moreover, the GCH model underestimates the near wake recovery, resulting in a larger momentum deficit at $3D$ (Figure~\ref{fig:ProfNEP}b) across the rotor region.
However, further downstream, at $5D$ from turbine T2 (Figure~\ref{fig:ProfNEP}c), the wake recovery is well captured by the GCH model as the curve collapses over that obtained from the LES.
In the wake of turbine T3 (Figure~\ref{fig:ProfNEP}d-f), it is observed that the GCH model slightly overestimates the momentum deficit.
Importantly, the ML model ameliorates the prediction of the GCH model as it collapses on top of the high-fidelity LES results, marking excellent agreement in the near and far wake for both the upstream turbine and that operating in the waked condition.
\begin{figure*}[tb]
 \begin{center}
  \psfrag{alab}[][][0.8]{a)}
  \psfrag{blab}[][][0.8]{b)}
  \psfrag{clab}[][][0.8]{c)}
  \psfrag{xlab}[][][1.0]{$x/D$}
  \psfrag{ylab}[][][1.0]{$z/D$}
  \psfrag{t1}[][][0.5]{T2}
  \psfrag{t2}[][][0.5]{T1}
  \psfrag{t3}[][][0.5]{T3}
  \psfrag{cblab}[][][1.0]{$\overline{U}/U_{hub}$}
  \includegraphics[width=\textwidth]{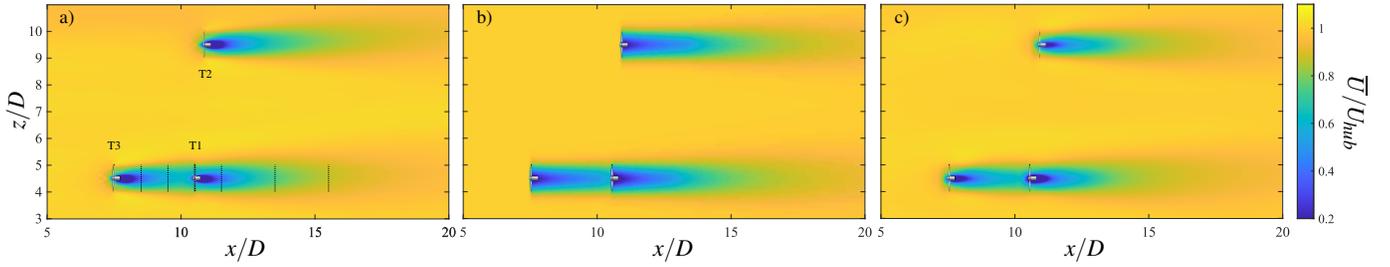}
  \caption{Contours of time-averaged velocity ($\overline{U}$) at a hub height plane for the West wind direction and hub height wind speed of $U_{hub}=6.1$ m/s at the SWiFT turbines facility; (a) LES, (b) GCH model, and (c) ML model. The streamwise and spanwise directions are given by $x/D$ and $y/D$, respectively, where $D$ is the turbine rotor diameter.}
  \label{fig:VelW}
 \end{center}
\end{figure*}

Further, contours of the time-averaged wind speed of the case of west wind direction at $U_{hub}=6.1$ m/s are shown in Figure~\ref{fig:VelW}. 
As seen, the wake flow of the turbines is similar to that of the case with the northeast wind direction.
However, due to the change in the wind direction, the turbines $T1$ and $T3$ are $3D$ apart.
The high momentum deficit observed in the near-wake of turbine $T3$ obtained from the GCH model impinges on the downstream turbine $T3$ (Figure~\ref{fig:VelW}b). 
As a result, this could lead to an underestimation of the power production relative to the LES (Figure~\ref{fig:VelW}a).
The wake flows of the turbines obtained from the ML model (Figure~\ref{fig:VelW}c) marks yet again a close resemblance to the LES results, which may significantly improve the power production prediction of the turbine downstream ($T1$).
\begin{figure}[tb]
 \begin{center}
  \psfrag{xlab}[][][1.0]{$\overline{U}/U_{hub}$}
  \psfrag{ylab}[][][1.0]{$(z-z_{T3})/D$}
  \psfrag{alab}[][][1.0]{a)}
  \psfrag{blab}[][][1.0]{b)}
  \psfrag{clab}[][][1.0]{c)}
  \psfrag{dlab}[][][1.0]{d)}
  \psfrag{elab}[][][1.0]{e)}
  \psfrag{elab2}[][][1.0]{f)}
  \includegraphics[width=0.48\textwidth]{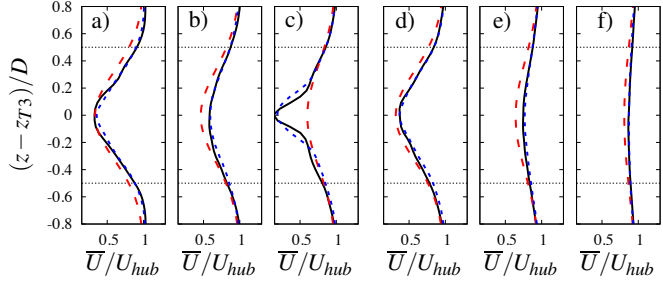}
  \caption{Time-averaged velocity ($\overline{U}$) profiles along the spanwise direction at a hub-height plane of the West wind direction $U_{hub}=6.1$ m/s case at (a) $1D$, (b) $2D$, (c) $3D$ downstream from turbine $T3$ and (d) $1D$, (e) $2D$, (f) $4D$ downstream from turbine $T1$; (\solid) LES, ({\color{red}\dashed}) GCH model and ({\color{blue}\dasheda}) ML model. The profiles were centered along the spanwise direction with the location of turbine $T3$ ($z_{T3}$).}
  \label{fig:ProfW}
 \end{center}
\end{figure}

Velocity profiles of the west wind direction and hub height wind speed of $U_{hub}=6.1$ m/s are depicted in Figure~\ref{fig:ProfW}.
Differences in the mean velocity in the near-wake of turbine $T3$ (Figure~\ref{fig:ProfW}a,b) are consistent with that observed for the northeast wind direction.
The velocity profile at $3D$ downstream from turbine $T3$ (Figure~\ref{fig:ProfW}c) corresponds to the velocity at the rotor of the turbine $T1$ downstream.
It should be noted that the difference in the wind speed obtained from the LES and the GCH model is caused by the presence of the nacelle that is modeled in the LES using the actuator surface model, while the GCH model neglects the nacelle.
Nevertheless, the ML model does capture the presence of the nacelle and, thus, shows better agreement than the GCH model.
In the wake of turbine $T1$ (Figure~\ref{fig:ProfW}d-e), the GCH model overestimates the momentum deficit compared to the LES results. 
The GCH model overestimation of the wake recovery for the west wind seems to be more pronounced than that of the northeast wind direction, which may be due to the proximity of the turbines. 
As seen, the near wake obtained from the GCH model is overestimated and may not fully recover when it impinges upon the downstream turbine. 
However, the ML model captures the velocity deficit in the wake of the turbine $T1$ and $T3$ markedly well, showing an excellent agreement with the numerical results from the LES.

\begin{figure*}[tb]
 \begin{center}
  \psfrag{alab}[][][0.8]{a)}
  \psfrag{clab}[][][0.8]{b)}
  \psfrag{xlab}[][][1.0]{$x/D$}
  \psfrag{ylab}[][][1.0]{$z/D$}
  \psfrag{cblab}[][][1.0]{$\delta$}
  \includegraphics[width=\textwidth]{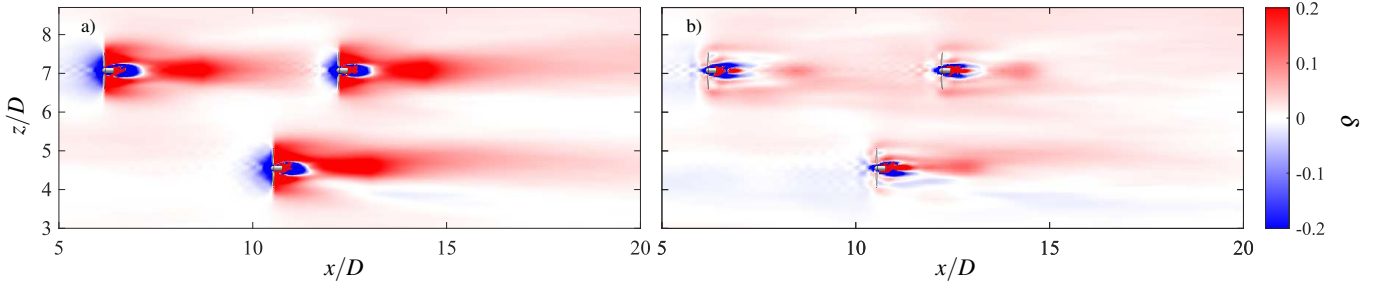}
  \caption{Contours of relative error ($\delta$) of the time-averaged velocity between that obtained from the (a) LES and GCH model and (b) LES and the ML enhanced model for a hub height wind speed of $U_{hub}=8.8$ m/s and North-East wind direction.}
  \label{fig:RE_NE}
 \end{center}
\end{figure*}
\begin{figure*}[tb]
 \begin{center}
  \psfrag{alab}[][][0.8]{a)}
  \psfrag{clab}[][][0.8]{b)}
  \psfrag{xlab}[][][1.0]{$x/D$}
  \psfrag{ylab}[][][1.0]{$z/D$}
  \psfrag{cblab}[][][1.0]{$\delta$}
  \includegraphics[width=\textwidth]{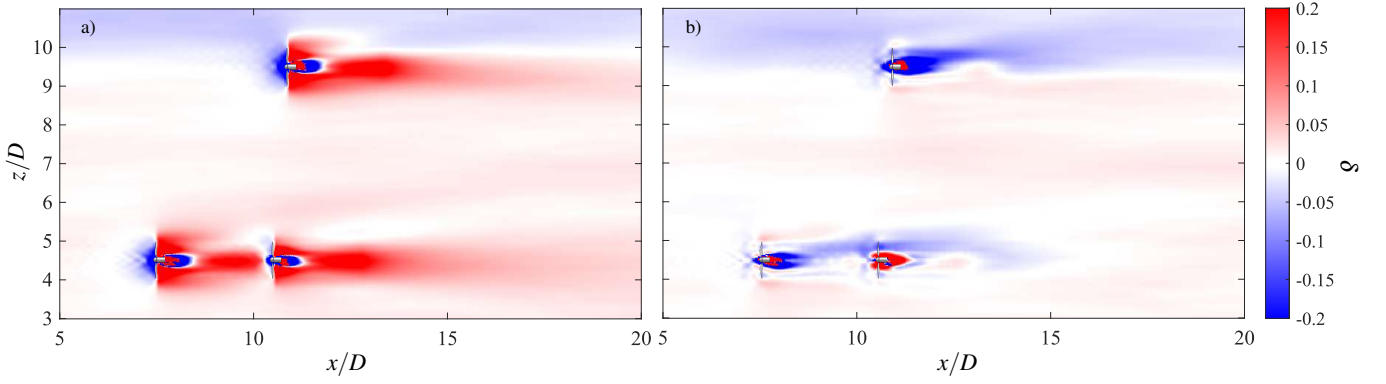}
  \caption{Contours of relative error ($\delta$) of the mean velocity between that obtained from the (a) LES and GCH model and (b) LES and the ML model for the hub height wind speed of $U_{hub}=6.1$ m/s and the west wind direction.}
  \label{fig:RE_W}
 \end{center}
\end{figure*}

The difference in the mean velocity field obtained from the LES and the wake models (i.e., GCH model and present ML model) were quantified by computing the relative error ($\delta$), as follows
\begin{equation}
    \delta = \frac{\overline{U}_{LES} - \overline{U}_{WM}}{ \overline{U}_{LES} },
\end{equation}
where $\overline{U}_{LES}$ and $\overline{U}_{WM}$ are the velocity obtained from the LES and the wake model, respectively.
Contours of the differential velocity $ \delta$ on a horizontal plane at hub height are shown in Figure~\ref{fig:RE_NE} and Figure~\ref{fig:RE_W} for the northeast and west wind directions, respectively. 
As seen in these figures, and also observed from the velocity profiles (Figures~\ref{fig:ProfNEP} and \ref{fig:ProfW}), the GCH model overestimates the momentum deficit in the near-wake region (Figures~\ref{fig:RE_NE}a and ~\ref{fig:RE_W}a) with a relative error of approximately $20\%$ in the center of the wake.
Although the error decreases to less than 10\% in the far wake ($(x-x_T)/D>4$), this low momentum region reaches the turbine located downstream for the west wind direction (Figure~\ref{fig:RE_W}a). 
Given that the power is proportional to the cube of the velocity ($P\propto U^3$), this can lead to a significant underestimation of the power production. 
As seen in Figures~\ref{fig:RE_NE}b and \ref{fig:RE_W}b, the present ML model also shows some deviation from the LES results ($|\delta|\leq 10\%$) in the near wake immediately behind the nacelle ($(x-x_T)/D<1$). 
It is argued that this is due to the complex dynamics of the vortex shedding at the nacelle, which the ML model has not fully captured. 
However, it should be noted that the velocity field obtained from the LES in this region is near zero, making a small error relatively larger.
Nevertheless, the ML model improves upon the results obtained from the GCH model, decreasing the relative error to $|\delta|<5\%$ in the near and far wake region of the turbines (Figures~\ref{fig:RE_NE}b and ~\ref{fig:RE_W}b).
\begin{figure}[tb]
 \begin{center}
  \psfrag{blab}[][][0.9]{a)}
  \psfrag{xlab}[][][0.7]{$(x-x_T)/D$}
  \psfrag{ylab}[][][0.6]{$\langle U_\mathrm{RA} \rangle/U_0$}
  \psfrag{cblab}[][][1.0]{$\delta$}
  \includegraphics[width=0.5\textwidth]{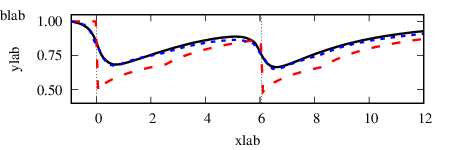}
  \psfrag{blab}[][][0.9]{b)}
  \includegraphics[width=0.5\textwidth]{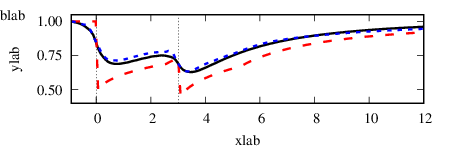}
  \caption{Rotor-averaged streamwise velocity ($\langle  U_\mathrm{RA} \rangle$)  in the streamwise direction and around the aligned pair of turbines of the (a) northeast wind direction $U_{hub}=8.8$ m/s case and (b) west wind direction $U_{hub}=6.1$ m/s case; (\solid) LES, ({\color{red}\dashed}) GCH model and ({\color{blue}\dasheda}) ML model. The rotor averaged velocity streamwise velocity $\langle  U_\mathrm{RA} \rangle$) is non-dimensionalized with the rotor averaged velocity obtained at $1D$ upstream from the turbine rotor ($U_0$) and the streamwise coordinate ($x$) is centered with respect to the rotor center ($x_T$).
   }
  \label{fig:RA}
 \end{center}
\end{figure}

To quantify the three-dimensional wake recovery, the time-averaged velocity over the rotor-swept area along the streamwise direction was computed as follows:
\begin{equation}
    \langle U_\mathrm{RA}\rangle = \frac{1 }{A_\mathrm{rot}} \int_{A_\mathrm{rot}} \overline{U} dA,
\end{equation}
where $\langle U_\mathrm{RA}\rangle$ is the momentum deficit over the rotor area, $U$ is the time-averaged streamwise velocity, and $A_\mathrm{rot}$ is the rotor area.
Figure~\ref{fig:RA} shows the rotor-averaged streamwise velocity in the streamwise directions and around the aligned pair of turbines for the northeast and west wind directions obtained from LES, ML, and GCH models. 
As demonstrated, the GCH model overpredicts the momentum deficit, leading to a smaller velocity over the rotor region relative to the LES results. 
It is also observed that the GCH model explicitly partitions the near and the far wake regions, showing a slower wake recovery---represented by the slope of $\langle U_\mathrm{RA}\rangle$---in the near wake and an increase in the wake recovery at around $(x-x_T)/D \approx 3$ for the far wake.
However, the rotor averaged velocity obtained from the ML model overlaps that of the LES.
Thus, one can say that the ML model has significantly augmented the fidelity of the results obtained from the GCH model by approximating the wake velocity at the downstream turbine with excellent accuracy in relation to the LES results.
\FloatBarrier

\subsection{Wake steering}
\begin{table}[b]
\caption{Root mean square error percentage of velocity field predictions with the GCH and ML model compared to the LES results for the wake steering cases. Empty spaces (---) indicate the training case.}
\centering
\resizebox{0.5\textwidth}{!}{
\begin{tabular}{lcccccccc cccccc}
\hline \hline
\multicolumn{1}{c}{$U_{hub}$ [m/s]} &     \multicolumn{14}{c}{Yaw Angle ($\gamma_{T1}$) }                                                                 \\
\hline
\multicolumn{1}{c}{}& \multicolumn{2}{c}{$-20^\circ$} & \multicolumn{2}{c}{$-15^\circ$} & \multicolumn{2}{c}{$-10^\circ$} & \multicolumn{2}{c}{$0^\circ$} & \multicolumn{2}{c}{$10^\circ$} & \multicolumn{2}{c}{$15^\circ$} & \multicolumn{2}{c}{$20^\circ$} \\
         & GCH       & ML        & GCH       & ML        & GCH        & ML          & GCH         & ML      & GCH       & ML       & GCH      & ML       & GCH     & ML   \\
5.4      & ---       & ---       & $8\%$     & $2\%$     & $8\%$      & $2\%$       & $8\%$       & $2\%$   & $8\%$     & $2\%$    & $8\%$    & $2\%$    & $8\%$   & $2\%$ \\
7.0      & $8\%$     & $2\%$     & $8\%$     & $2\%$     & $8\%$      & $2\%$       & $8\%$       & $2\%$   & $8\%$     & $2\%$    & $8\%$    & $2\%$    & $8\%$   & $2\%$ \\
8.5      & $8\%$     & $2\%$     & $8\%$     & $2\%$     & $8\%$      & $2\%$       & ---         & ---     & $8\%$     & $2\%$    & $8\%$    & $2\%$    & $8\%$   & $2\%$ \\
10.1     & $8\%$     & $2\%$     & $8\%$     & $2\%$     &   ---      & ---         & $6\%$       & $3\%$   & $8\%$     & $2\%$    & $8\%$    & $2\%$    & $8\%$   & $2\%$ \\
11.6     & $8\%$     & $2\%$     & $8\%$     & $2\%$     & $8\%$      & $2\%$       & $7\%$       & $3\%$   & $8\%$     & $2\%$    & $8\%$    & $2\%$    & ---     & ---   \\
\hline \hline
\end{tabular}
}
\label{table:rmse_yaw}
\end{table}

The training of the ML model for wake steering was performed using the cases: $U_{hub}=5.4$ m/s with $\gamma_{T1} =-20^\circ$, $U_{hub}=8.5$ m/s with $\gamma_{T1} =0^\circ$,  $U_{hub}=10.1$ m/s $\gamma_{T1} =-10^\circ$ and  $U_{hub}=11.6$ m/s $\gamma_{T1} =20^\circ$ (Figure~\ref{fig:CB} marks the relative positioning of yawed turbine $T1$).
It is important to highlight that the selection of training cases was done deliberately to ensure that the ML comprehensively captures the diverse conditions under which the turbines operate in an interpolation manner. 
While the ML model might offer a reasonable representation of wake characteristics for extrapolation cases (i.e., $\gamma_{T1}>20^\circ$), no modeling has been extended beyond the parameters used in training.
Hence, the subsequent analysis is focused on \textit{interpolation} cases.
\begin{figure*}[tbh]
 \begin{center}
  \psfrag{alab}[][][0.7]{a)}
  \psfrag{blab}[][][0.7]{b)}
  \psfrag{clab}[][][0.7]{c)}
  \psfrag{xlab}[][][0.8]{ }
  \psfrag{ylab}[][][0.7]{$z/D$}
  \psfrag{cblab}[l][b][0.5][180]{$\overline{U}/U_{hub}$}
  \includegraphics[width=\textwidth]{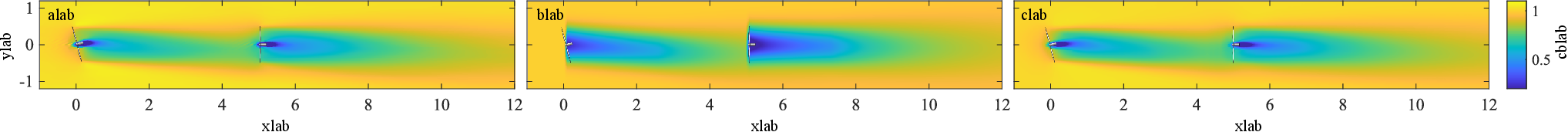}
  \psfrag{alab}[][][0.7]{d)}
  \psfrag{blab}[][][0.7]{e)}
  \psfrag{clab}[][][0.7]{f)}
  \psfrag{xlab}[][][0.7]{$x/D$}
  \includegraphics[width=\textwidth]{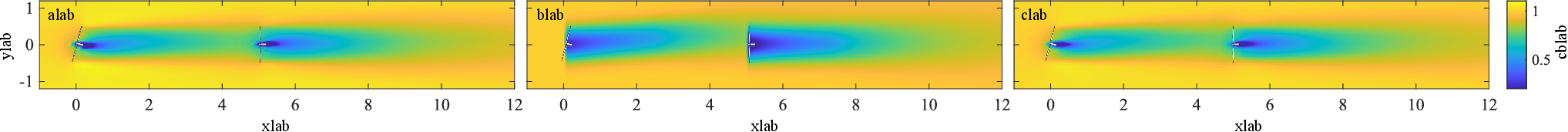}
  \caption{Contours of time-averaged velocity ($\overline{U}$) at a hub height plane of the aligned wind turbines of the (a-c) $U_{hub}=10.1$ m/s with $\gamma_{T1}=15^\circ$ and (d-f) $U_{hub}=7.0$ m/s with $\gamma_{T1}=-15^\circ$ cases; (a,d) LES, (b,e) GCH model, and (c,d) ML model. 
  }
  \label{fig:VelYaw}
 \end{center}
\end{figure*}

The root mean square error of the wind speed in the SWiFT facility obtained from the GCH and ML model are summarized in Table~\ref{table:rmse_yaw}. 
As presented in this table, the GCH model consistently obtained root mean square errors that range between $6\%$ and $8\%$, while the ML model decreases the error to the range of $2\%$ to $3\%$. 
For brevity, the analysis of the velocity field with wake steering is conducted for two specific cases of $U_{hub}=10.1$ m/s with $\gamma_{T1} =15^\circ$ and $U_{hub}=7.0$ m/s with $\gamma_{T1} =-15^\circ$. 
The former involves a yaw angle not included in the training parameters ($\gamma_{T1} =15^\circ$), while the latter encompasses both a wind speed ($U_{hub}=7.0$ m/s) and a yaw angle ($\gamma_{T1} =-15^\circ$) that were absent from the training dataset.
Despite the difference in their wind speed ($U_{hub}$), for the sake of brevity, they will be referred to by their respective yaw angles ($\gamma_{T1} = 15^\circ$ and $\gamma_{T1} =-15^\circ$).
\begin{figure}[tb]
 \begin{center}
  \psfrag{L1}[][][1.0]{a)}
  \psfrag{ylab}[][][1.0]{$z/D$}
  \psfrag{xlab}[][][1.0]{}
  \includegraphics[width=0.45\textwidth,trim={0 10 0 0},clip]{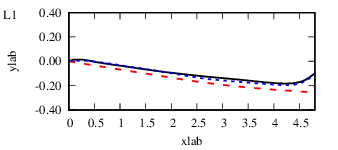}
  \psfrag{L1}[][][1.0]{b)}
  \includegraphics[width=0.45\textwidth,trim={0 10 0 2},clip]{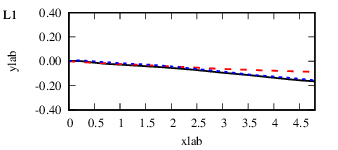}
  \psfrag{L1}[][][1.0]{c)}
  \includegraphics[width=0.45\textwidth,trim={0 10 0 2},clip]{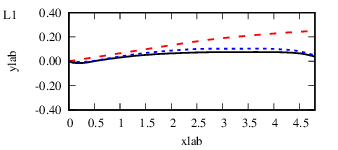}
  \psfrag{L1}[][][1.0]{d)}
  \psfrag{xlab}[][][1.0]{$x/D$}
  \includegraphics[width=0.45\textwidth,trim={0 0 0 2},clip]{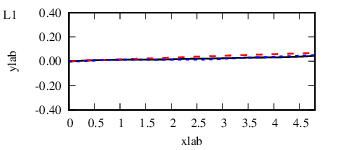}
  \caption{
  Wake centerline of turbine (a,c) T1 and (b,d) T2 of the (a,b) $U_{hub}=10.1$ m/s, $\gamma_{T1}=15^\circ$ and (c,d) $U_{hub}=7.0$ m/s and $\gamma_{T1}=-15^\circ$ cases; (\solid) LES, ({\color{red}\dashed}) GCH model and ({\color{blue}\dasheda}) ML model.
  }
  \label{fig:WakeCent}
 \end{center}
\end{figure}

Figure~\ref{fig:VelYaw} depicts the contours of the velocity field obtained from the LES, GCH, and ML model at a hub height plane. 
As observed in the velocity contours obtained from the LES (Figure~\ref{fig:VelYaw}a and d), the wake of the turbines is skewed toward the direction of the normal vector to the turbine rotor plane.
This phenomenon arises from the thrust force exerted by the turbine rotor on the flow, forming two counter-rotating vortices that displace the wake from its central position~\cite{fleming2018a}.
Although the GCH model does not show an angled rotor (Figure~\ref{fig:VelYaw}b and e), it redirects the wake caused by the yaw misalignment. 
Further, like the un-yawed cases, the GCH model overpredicted the momentum deficit in the near wake of both turbines ($T1$ and $T2$).
The wakes of the turbines obtained from the ML model (Figure~\ref{fig:VelYaw}c and f), however, show a greater similarity to that obtained from the LES.

The wake deflection of the ML and GCH models is examined by quantifying the trajectory of the wake centerlines using Gaussian fits to the momentum deficit at each computational grid node in the streamwise direction. 
The centerline location is then identified as the position of the maximum of the Gaussian fit at each computational node. 
The computed wake centerlines are shown in Figure~\ref{fig:WakeCent}.
Comparison of the wake centerline of the $\gamma_{T1}=15^\circ$ shows a slight overestimation of the wake deflection by the GCH model for the upstream turbine (Figure~\ref{fig:WakeCent}a).
Moreover, the secondary steering, caused by the heterogeneity of the velocity at the downstream turbine, obtained from the GCH model, shows good agreement with the LES in the near wake (Figure~\ref{fig:WakeCent}b). 
However, the GCH model yet again underpredicted the wake deflection in the far wake ($x/D>3$), marking a pronounced difference with the LES results at $x/D\approx5$.
We note that the LES wake centerline of the $\gamma_{T1}=-15^\circ$ case (Figure~\ref{fig:WakeCent}c) has a smaller deflection than for the positive yaw angle that is in agreement with the observations of Fleming et al.~\cite{fleming2018a}. 
Given that the GCH model lacks wake rotation, its wake deflection is symmetrical without differentiating between positive and negative yaw angles. 
Therefore, as seen, the wake centerline of the GCH model markedly deviates from that of LES.
The ML model successfully captures this asymmetrical deflection of the wakes, demonstrating a close agreement with the LES results. 
Nevertheless, the secondary steering predicted by the GCH model for the $\gamma_{T1}=-15^\circ$ case shows to agree with that of the LES.
This may be caused by the overprediction in the effective yaw angle at the downstream turbine, which coincidentally led to a better prediction for the wake deflection.

\section{Final Remarks}\label{sec:summary}
An ultra-efficient machine learning-based model has been developed to predict wind farms time-averaged wake flow fields.
This novel model successfully closes the computational cost divide between an engineering wake model and large eddy simulations, all while maintaining high-fidelity in the description of wind farm turbine wakes.
Leveraging an autoencoder convolutional neural network inspired by the proven U-Net architecture~\cite{ronneberger2015}, this model stands out for its remarkable speed and cost-effectiveness.
During the training process, the presented ML model uses low- and high-fidelity datasets as its input and target vectors, respectively. 
Notably, once trained, this outstanding machine learning model demonstrates its efficiency by seamlessly generating high-fidelity targets from solely low-fidelity input data.
This capability significantly reduces costs compared to traditional high-fidelity simulations, making it a groundbreaking and cost-effective solution for wind farm predictions.

To evaluate the performance of the proposed ML model, a series of large eddy simulations were performed to generate high-fidelity data for the wake flow field of the SWiFT facility. 
The simulations were performed using VFS-Wind, which solves the filtered Navier-Stokes equation in a curvilinear grid, and the wind turbine blades and nacelle were parameterized using the actuator surface model. 
In these simulations, hub-height wind speeds of $U_{hub} = 4.8$, $6.1$, $7.5$, $8.8$, and $10.2$ m/s with various wind directions of $30^\circ$ (northeast), $180^\circ$ (south), $210^\circ$ (southwest), and $266^\circ$ (west) were considered. 
Moreover, an additional set of simulations was conducted for the south wind direction configuration to assess the ML model performance for wake steering. 
We imposed fixed yaw angles of $\gamma_{T1} = -20^\circ$, $-15^\circ$, $-10^\circ$, $0^\circ$, $10^\circ$, $15^\circ$, and $20^\circ$ to the upstream turbine $T1$.

The input to the ML model consists of the three-dimensional mean velocity field prediction obtained from the low-fidelity Gauss Curl Hybrid model~\cite{King2021}. 
These were obtained from the FLORIS (v3.4) package~\cite{Fleming2020}.
The input parameters for this analytical wake model description were obtained from the example input file provided by the FLORIS (v3.4) package. 
Although fine-tuning these parameters is recommended for an accurate description of the turbine wake, they were used as is without any adjustment to ensure that the developed ML model is not sensitive to these parameters of the low-fidelity model. 
The ML model was trained for $50\,000$ epochs to minimize the mean square error between network outputs and LES simulations using the Adam optimizer~\cite{Kingma2015} with a learning rate of $10^{-4}$.

A comparative analysis of the time-averaged wake velocity field obtained from the LES, GCH, and ML models was performed. 
It was demonstrated that the kCH model overestimated the momentum deficit of the wake flows, particularly in the near wake region.
In contrast, the ML model produced a more precise representation of the wake flow with a relative error of less than $5\%$ in both the near and far wake of the turbines.
Furthermore, the rotor-averaged velocity obtained from the ML model aligned closely with that from the LES, suggesting a potential for more accurate downstream turbine power production predictions.
Notably, the ML model successfully captured the asymmetric wake deflection observed for opposing yaw angles, exhibiting excellent agreement with LES results. 
The proposed ML model improves the GCH model, which lacks consideration for the interaction of the rotating wake with counter-rotating vortex pairs resulting from yaw misalignment and results in a symmetric deflection.

Each LES run costs about $73\,000$ CPU hours, corresponding to the computational time required to produce the mean flow field of the SWiFT facility. The training of the ML model took around $100$ CPU hours.
The inputs to the ML model were obtained from the FLORIS (v3.4) package, which has a computational time of around $0.03$ CPU hours. 
After training, the ML model evaluation has a computational time of $0.1$ CPU hours.
Therefore, the total time required by the ML model is estimated to be around $0.13$ CPU hours. 
This reduction, exceeding $99\%$ compared to LES, underscores the capacity of the ML model to efficiently deliver a high-fidelity representation of turbine wakes. 
Therefore, it demonstrates the ability of the proposed ML model to obtain a high-fidelity description of the turbine wakes for control co-design optimization of utility-scale wind farms by integrating an engineering wake model with high-fidelity data.  

\begin{acknowledgments}
This research was supported by grants from the U.S. Department of Energy's Office of Energy
Efficiency and Renewable Energy (EERE) under the Water Power Technologies Office
(WPTO) Award Number DE-EE0009450, the National Offshore Wind Research and Development Consortium (NOWRDC) under agreement number 147503, and NSF (grant number 2233986). The views expressed herein do not necessarily represent the view of the U.S. Department of Energy or the United States Government.
\end{acknowledgments}

\section*{Data Availability Statement}
The Virtual Flow Simulator model~\cite{aksen_2024}, the machine learning model~\cite{khosronejad_2024}, and the three-dimensional velocity fields~\cite{khosronejad_2024b} in this work are made available in the online open repository of Zenodo.

\bibliography{references}

\end{document}